\documentclass{iopconfser}

\usepackage{graphicx}  
\usepackage{wrapfig}

\begin{document}

\title{Dark matter direct detection: status, results and future plans}

\author{Laura Baudis}

\affil{Physik-Institut, University of Z\"urich,  Winterthurerstrasse 190, 8057  Z\"urich, Switzerland}
\email{laura.baudis@uzh.ch}

\begin{abstract}

Direct dark matter detection experiments search for rare signals induced by hypothetical, galactic dark matter particles in low-background detectors operated deep underground. I will briefly review the direct detection principles, the expected signals and backgrounds, and the main experimental techniques. I will then discuss the status of ongoing experiments aiming to discover new particles in the keV - TeV mass range, as well as future detectors and their sensitivity.

\end{abstract}
\vspace{2pc}

\section{Introduction}

There is abundant evidence for dark matter (DM) across many time and  length scales in the Universe. It relies on measurements of galactic rotation curves and orbital velocities of individual galaxies within clusters, on cluster mass determinations via gravitational lensing and the distribution of the hot, X-ray emitting gas, on precise measurements of the cosmic microwave background temperature fluctuations and of the abundance of light elements, and upon the mapping of structures on the largest scales~\cite{Bertone:2004pz}. In addition, cold dark matter  is one of the foundations of the standard model of cosmology, $\Lambda$CDM, accounting for 26.4\% of the critical density, or 84.4\% of the total matter density~\cite{Planck:2018vyg}.  Notwithstanding, the DM is observed only indirectly, through its gravitational effects on luminous matter, and its fundamental nature remains a mystery.  An intriguing possibility is that DM is made of new elementary particles, yet to be identified. Such particles would carry no electric and colour charge, would have rather weak self-interactions and would be either stable or very long-lived. Various extensions of the Standard Model of particle physics provide viable candidates,  and searches can be classified as direct, indirect, and accelerator-based~\cite{ParticleDataGroup:2022pth}. Here I will discuss direct searches. After a brief review of DM candidates and of direct detection principles, I will review the expected signals and sources of backgrounds and the main experimental techniques. I will discuss recent results and detection prospects for this and the next decade.

\section{Dark matter candidates}

\begin{figure}[h!]
\centering
\includegraphics[width= 1.0\textwidth]{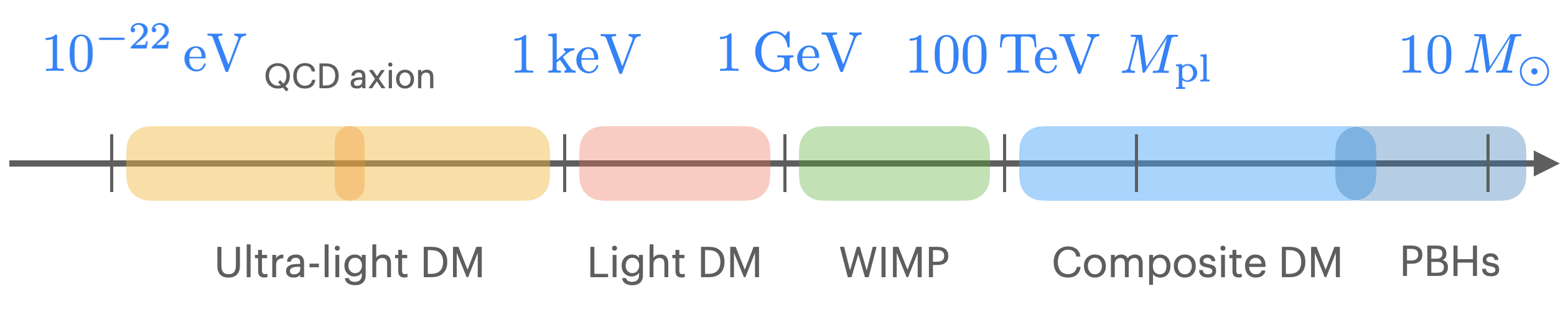}
\caption{The broadly allowed mass range for dark matter candidates (not to scale). Inspired by Figure\,3 from~\cite{Lin:2019uvt}.}
\label{fig:dm_mass_scale}
\end{figure}

Candidates for DM extend over a large range of masses and interaction cross sections. As shown in Figure~\ref{fig:dm_mass_scale}, the mass extends from about 10$^{-22}$\,eV, up to tens of solar masses. Two classes of models stand out: Weakly Interacting Massive Particles (WIMPs) and axions, for these are theoretically well-motivated by open questions in particle physics~\cite{ParticleDataGroup:2022pth}. Here I focus on light DM (sub-GeV masses) and WIMPs; the search for axions is covered by Andreas Ringwald in this volume.

\section{Principles of direct dark matter detection}

\begin{figure}[h!]
\includegraphics[width= 0.45\textwidth]{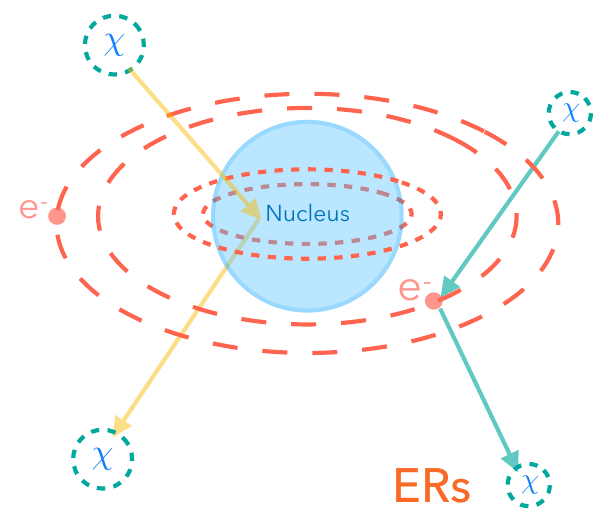}
\includegraphics[width= 0.5\textwidth]{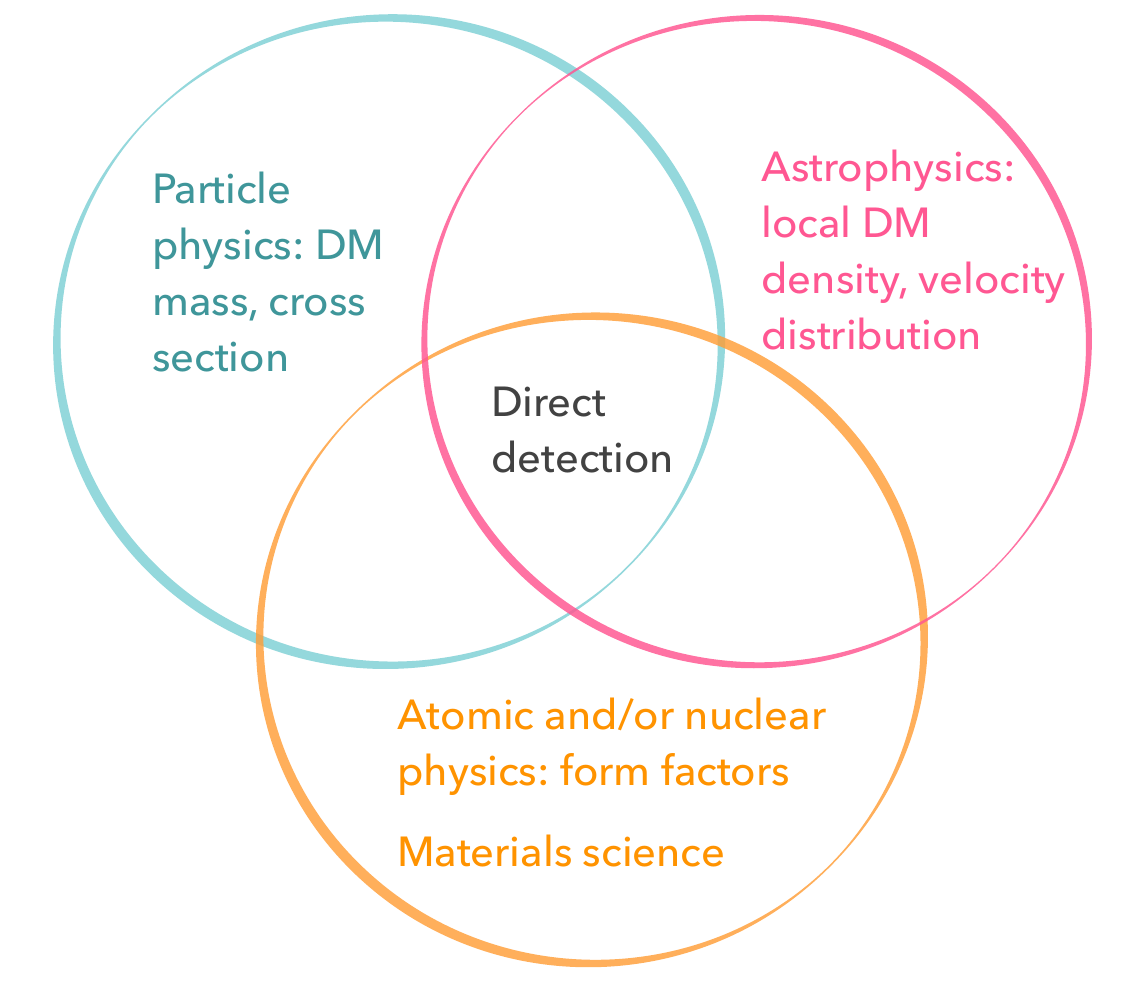}
\caption{(Left): Hypothetical dark matter particle candidates can scatter off atomic nuclei (NRs) or interact with the electrons in the atomic shell (ERs), via inelastic scattering or absorption. (Right): Direct dark matter detection requires input from several research fields. } 
\label{fig:er_nr}
\end{figure}

Direct DM detection experiments search for elastic or inelastic scatters of particle candidates with atomic nuclei or with electrons in the atomic shells of various target materials. In general, the former are classified as nuclear recoils (NRs), and the later as electronic recoils (ERs), as shown schematically in Figure~\ref{fig:er_nr}, left. The main physical observable is a differential recoil spectrum (or a line feature for the absorption of, e.g.,  axions and dark photons), and its modelling relies on several  inputs from particle physics (particle mass and interaction cross section), astrophysics (local DM density and velocity distribution, galactic escape velocity), atomic and nuclear physics (form factors), as well as material science, for interactions do not occur on free atoms or electrons.

If DM particles scatter off atomic nuclei, the differential scattering rate $R$ as a function of nuclear recoil energy $E_R$ can be written as:

\begin{equation}
\frac{dR(E_R,t)}{dE_R} = N \frac{\rho_0}{m_{\chi}}  \int_{v > v_{min}} v f(\vec{v} + \vec{v}_E(t)) \frac{d\sigma (E_R,v)}{dE_R} d^3 v,
\end{equation}

\noindent
where $N$ is the number of target nuclei, $m_{\chi}$ is the mass of the DM particles, $v = |\vec{v}|$ is the speed of the particle in the experiment's rest frame, $ f(\vec{v} + \vec{v}_E(t))$ is the velocity distribution in the Earth's frame, $v_{min}$ is the minimum speed of the DM particles that can cause a recoil  energy $E_R$ and $\sigma$ is the scattering cross section on the  nucleus~\cite{Jungman:1995df,Lewin:1995rx}. For elastic scattering, the minimum  velocity is $v_{min} = (m_N E_R/2\mu^2)^{1/2}$, with $m_N$ being the mass of the 
nucleus, and $\mu=(m_N m_{\chi})/(m_N + m_{\chi})$ the reduced mass of the nucleus-DM system. For inelastic scattering, the minimum speed 
becomes $v_{min} = (m_N E_R/2\mu^2)^{1/2} + E^*/(2 m_N E_R)^{1/2}$, with  $E^*$ being the nuclear excitation energy. The prompt de-excitation energy, if observed in addition to $E_R$, will boost the region-of-interest to higher energies and allows for the determination of the mass of the DM particle~\cite{Ellis:1988nb,Baudis:2013bba}. 

For light DM particles ($m_{\chi} <1$\,GeV) searches for DM-nucleus scattering rapidly loose sensitivity, due to energy thresholds around a few 100\,eV - few keV. Direct detection experiments thus search for inelastic scatters off bound electrons, allowing for all of the kinetic energy  to be transferred to the material~\cite{Essig:2011nj}. Some of the possibilities are ionisation, excitation, and molecular dissociation processes, which  require energies in the range (1-10)\,eV, allowing to probe DM masses down to the $\mathcal{O}$(MeV) range, or even below.  The signal depends on the material, and can consist of one or more electrons (in semiconductors, noble liquids, graphene), one or more photons  (in scintillators) or phonons (in superconductors and superfluids) and quasiparticles (in superconductors).   Predicted differential rates in various materials can be found, e.g., in ~\cite{Essig:2011nj, Catena:2022fnk,Trickle:2022fwt}. 

\section{Signals and experimental techniques}

\begin{figure}[h!]
\centering
\includegraphics[width= 0.8\textwidth]{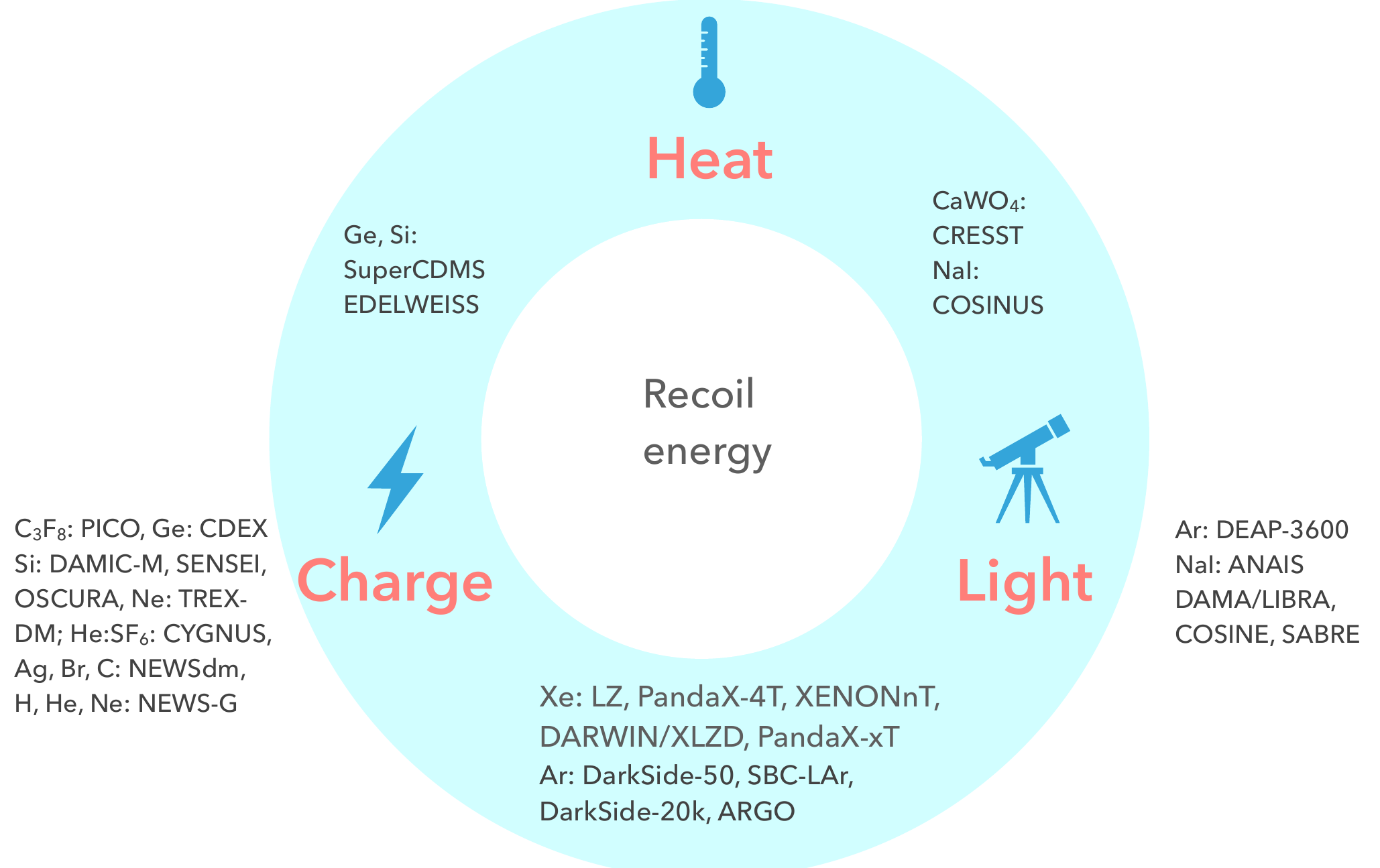}
\caption{Schematic view of the main experimental techniques employed in direct DM detection, together with ongoing or proposed experiments (not a complete list).}
\label{fig:techniques}
\end{figure}

Direct DM detection experiments are designed to observe low-energy (keV-scale and below)  and rare  (fewer than $\sim$1\,event/(kg\,y)) signals which are induced by DM particle scatters  in a detector operated deep underground.  The observed signals are mostly in the form of ionisation, scintillation or lattice vibrations, as shown schematically in Figure~\ref{fig:techniques}. Many experiments detect more than one signal, which allows one to distinguish between ERs and NRs. A three dimensional position resolution is required to define central  detector regions (called fiducial volumes) with low background rates due to the radioactivity of surrounding detector components. The ability to separate single- and multiple-scatters is used to reject a significant fraction of backgrounds, considering that a DM particle will scatter at most once in a given detector. 

\section{Backgrounds}

Historically, the main backgrounds in direct detection experiments were due to the radioactivity of detector materials and surroundings, to cosmic muons and their secondaries, as well as  cosmic activation of detector components, including the DM target~\cite{Heusser:1995wd}. These are strongly reduced by selecting radio-pure materials, by going deep underground and surrounding the detectors with active neutron and muon vetos, and by minimising the exposure to cosmic rays at the Earth's surface. While many of these background sources are still relevant today, and must be further suppressed for next-generation experiments, the final sources will be astrophysical neutrinos, which are impossible to shield~\cite{Billard:2013qya}. 

The neutrino-induced backgrounds, in particular those from solar neutrinos, can be mitigated to some extend by detectors with directional information~\cite{Grothaus:2014hja,OHare:2015utx}.  On the other hand, a measurement of various neutrino fluxes provides exciting science opportunities for DM detectors. These will be sensitive to astrophysical neutrinos from the Sun, from supernovae and from the atmosphere. I refer to Ref.~\cite{Dutta:2019oaj} for a review on the topic.

Apart from events due to particle interactions, instrumental effects or artefacts can also play a significant role, in particular at lowest energies close to the energy threshold of a detector. Examples are the low-energy excess observed in bolometers~\cite{Fuss:2022fxe} and the combinatorial (or accidental)  background in noble liquid detectors~\cite{XENON:2020gfr,LZ:2022ysc}.

\section{Experiments}
 
\begin{figure}[h!]
\includegraphics[width= 0.47\textwidth]{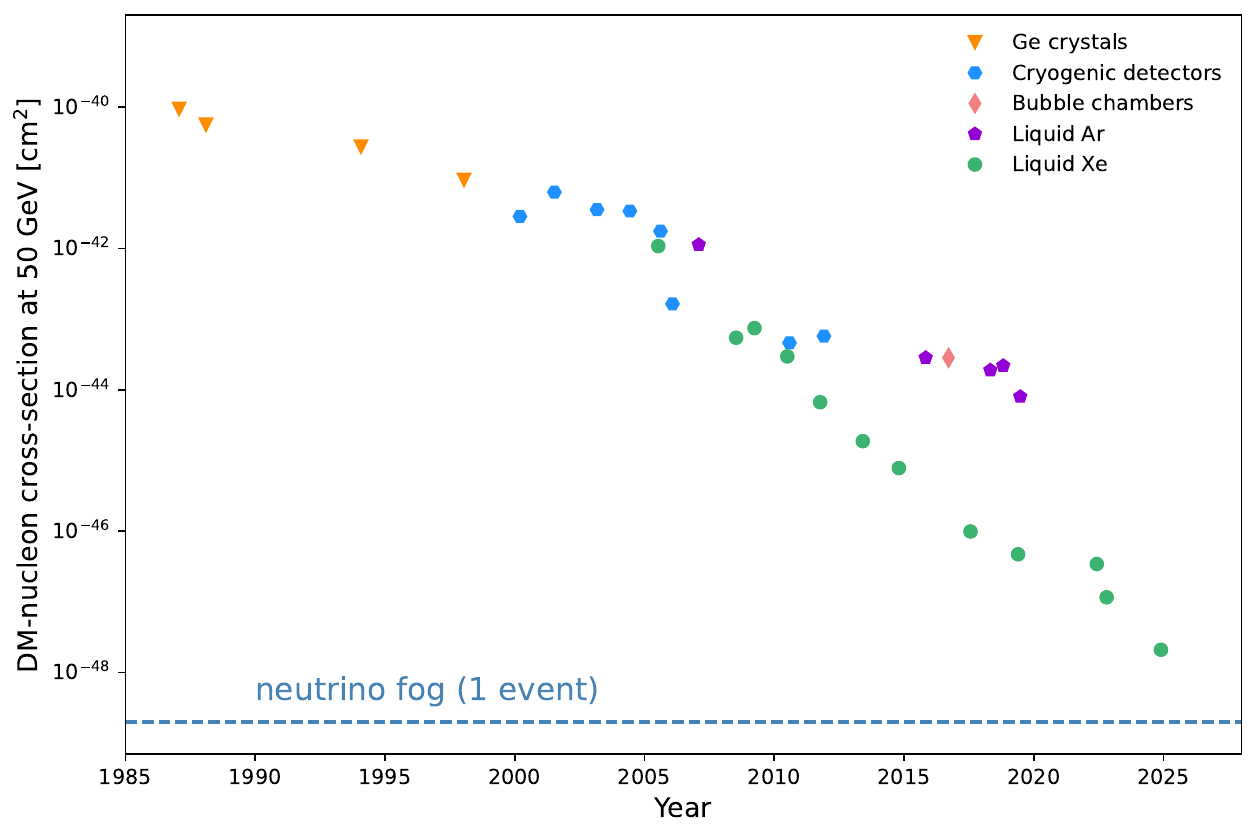}
\includegraphics[width= 0.52\textwidth]{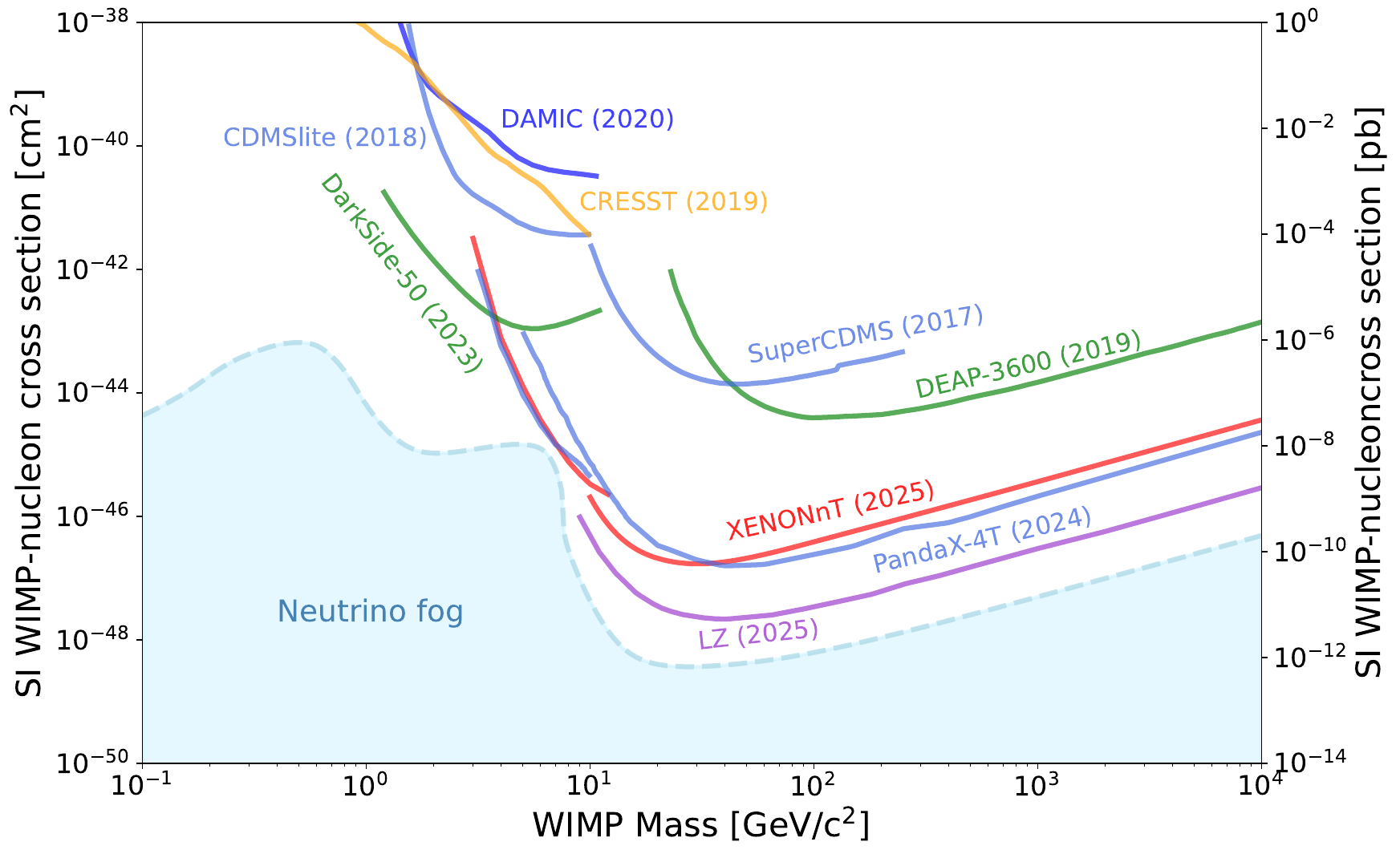}
\caption{(Left):  Time evolution of upper limits on the SI WIMP-nucleon cross section for a 50\,GeV WIMP. (Right): Upper limits on the SI WIMP-nucleon cross section as a function of DM mass. The region where astrophysical neutrinos will start to be a limiting background is indicated in blue (neutrino fog). Figure from~\cite{ParticleDataGroup:2024cfk}, updated (PDG2025).}
\label{fig:limits}
\end{figure}

Many  techniques are employed to search for the rare interactions expected from galactic DM particles. While most experiments have sensitivity to a wide range of particle masses, some technologies were optimised for light and heavy DM, respectively. There is an ongoing R\&D programme to observe the direction and sense of a nuclear recoil, correlated with the direction of the incoming WIMP.  The time evolution of upper limits on the spin-independent (SI) WIMP-nucleon cross section for a 50\,GeV WIMP is shown in Figure~\ref{fig:limits}, left, while  the right panel shows current constraints as a function of WIMP mass.  In the following I will briefly discuss some of the main direct detection techniques, as well as ongoing or proposed experiments. 
 
\subsection{Cryogenic detectors}

Solid-state detectors operated at sub-Kelvin temperatures can reach low energy thresholds, excellent energy resolution and are very effective in distinguishing NRs from ERs on an event-by-event basis.   Because of the T$^3$-dependence of the heat capacity of a dielectric crystal,  at low temperatures a small energy deposition can significantly change the temperature of the absorber~\cite{Booth:1996xdy}. The  temperature increase is measured with different types of temperature sensors: athermal sensors measure the fast, non-equilibrium phonons, while thermal sensors measure the phonons after they thermalise. Examples for athermal and thermals sensors are transition edge sensors (TES) and neutron transmutation doped germanium sensors (NTDs), respectively. If the crystal scintillates or is a semiconductor, an additional signal, light or charge, can be detected and employed for particle discrimination.

Dark matter experiments based on the bolometric technique are SuperCDMS~\cite{Zatschler:2024ssq}, EDELWEISS~\cite{Guy:2024xog} and CRESST~\cite{Banik:2024kwt}. SuperCDMS uses Ge and Si crystals with TES readout, in two detector types, high voltage (HV) and interleaved Z-dependent ionisation and phonon (iZIP) detectors. The iZIP detectors measure both phonon and ionisation signals, allowing one to discriminate between NR and ERs, while the HV detectors measure the phonon signal only.  A bias voltage of 100\,V is applied to the HV crystals, which amplifies the primary ionisation signal via the Neganov-Trofimov-Luke (NTL) effect, resulting in a lower energy threshold. The collaboration is currently  constructing a new facility at SNOLAB, with an initial payload of 18 Ge and 6 Si detectors, for a total mass of 30\,kg. The detectors will be configured into four towers, each containing six crystals with 1.4\,kg of mass for Ge and 0.6\,kg for Si crystals. Two towers will house detectors operated in HV mode, optimised for WIMP masses down to 0.3\,GeV. The remaining detectors will be operated in iZIP mode.

EDELWEISS, located at  Laboratoire Souterrain de Modane (LSM), uses 24 Ge crystals,  860\,g each,  with NTD Ge sensors for phonon readout, as well as concentric aluminium ionisation sensors interleaved on all absorber surfaces. This allows for fiducialisation and thus separation between bulk and surface events. Some of the detectors are operated with NTL amplification to achieve a lower energy threshold, while others are equipped with NbSi TES, to probe the origin of "heat only" (hence no charge detected) events. A new development in the collaboration is  CRYOSEL, which operates smaller, 40\,g Ge detectors with NTDs and small superconducting single electron devices (SSED) to tag single charges produced in the crystals. The electrodes for ionisation readout are made of Al, while the SSEDs are made of a lithographed NbSi line. These detectors are optimised for light DM searches.

CRESST, operated at Laboratori Nazionali del Gran Sasso (LNGS), employs a variety of scintillating crystals (CaWO$_4$, LiAlO$_2$, Al$_2$O$_3$, and Si), including ten 25\,g CaWO$_4$ crystals with tungsten TES readout. Since the crystals scintillate, they are also equipped with silicon-on-sapphire light detectors with TES readout, allowing one to separate NRs and ERs. The crystal housings are scintillating as well, and their holders are instrumented with TES, features which are used to veto surface-backgrounds, as well as signals from particles interacting in surrounding materials. CRESST also operates new diamond crystals with masses of 0.175\,g and energy thresholds for NRs of 17\,eV.  These show high potential for sub-GeV DM particles.

A common feature to all solid state experiments operated at low temperatures is a sharply rising energy spectrum below 200\,eV, of unknown origin.  This triggered a large community effort and the EXCESS workshop series~\cite{Fuss:2022fxe}, with the goal to understand and mitigate this new source of backgrounds which limits the sensitivity to light DM. A likely cause are long-lived metastable states releasing energy in small bursts in the systems.

\subsection{Noble liquids}

Detectors using liquefied noble gases are non-segmented, homogeneous, compact and self-shielding.  In single-phase configuration, the scintillation light is observed with photosensors surrounding the liquid volume. These feature a high light yield and a relatively simple geometry, with no applied electric fields.  In two-phase (liquid and gas) time projection chambers (TPCs), the simultaneous detection of ionisation and scintillation signals allows for discrimination between ERs and NRs and for the determination of the three dimensional  position of an interaction.
Liquid argon (LAr) and liquid xenon (LXe) are employed as DM targets, while R\&D on liquid helium (LHe) is ongoing.  LAr experiments also employ a powerful  pulse shape discrimination technique for effective separation between ER and NR events, at the expense of higher energy thresholds  than in LXe detectors~\cite{ParticleDataGroup:2022pth}.

Single-phase LAr and LXe detectors are DEAP-3600~\cite{Viel:2024qjs} at SNOLAB and XMASS~\cite{XMASS:2022tkr} at Kamioka, respectively. XMASS took data until 2019, while DEAP-3600 is being upgraded, with a new run to start in 2025. DEAP-3600, with 3.3\,t (1\,t) of LAr in total (fiducial)  in a large ultralow-background acrylic cryostat, recently set leading constraints on Planck-scale DM~\cite{DEAPCollaboration:2021raj}. Two-phase Ar and Xe detectors are DarkSide-50~\cite{Agnes:2024dms} at LNGS (50 kg of LAr in total, depleted in $^{39}$Ar,  data until 2019) and LUX-ZEPLIN (LZ)~\cite{LZ:2019sgr} at the Sanford Underground Facility (SURF), PandaX-4T~\cite{PandaX-4T:2021bab} at the China Jing Ping Laboratory (CJPL) and XENONnT ~\cite{XENON:2024wpa} at LNGS, all operating multi-tonne Xe TPCs. DarkSide-50 set leading exclusion limits for WIMP masses of a few GeV. The three large LXe experiments presented first results on WIMPs and other DM candidates from their early science runs, and continue to acquire data towards their design exposures and sensitivities.

Next-generation detectors are DarkSide-20k~\cite{Zani:2024ybb}, in construction phase, and Argo, DARWIN/XLZD~\cite{Aalbers:2016jon,Baudis:2024jnk}, PandaX-xT~\cite{PandaX:2024oxq},  in design and R\&D phase. DarkSide-20k will house 51\,t of underground LAr, with 20\,t in the fiducial volume, in an octogonal TPC viewed by SiPM arrays. The cryostat is being constructed at LNGS, with the first detector filling expected for the end of 2026. Argo will be a much larger, 400\,t LAr detector, likely located at SNOLAB. Assuming a 200\,t\,y exposure, DarkSide-20k will have a 5-$\sigma$ discovery sensitivity for 1\,TeV WIMPs with a SI cross section on nucleons of $2.1 \times 10^{-47}$\,cm$^2$. 
DARWIN would operate a TPC with 40\,t of LXe in the active region (50\,t in total)~\cite{Aalbers:2016jon}. In 2021 the LZ, XENON and DARWIN collaborations signed an  MoU and joined forces to form the XLZD consortium, with the goal of constructing and operating the next-generation LXe experiment together.  The size and scope of the detector will be enlarged, compared to DARWIN, with a cylindrical 3\,m$\times$3\,m TPC containing 60\,t of LXe (75\,t in total). In this baseline design, the  TPC is placed in a low-background, double-walled cryostat surrounded by neutron and muon vetos. 
XLZD would allow for a 3-$\sigma$ WIMP discovery at a SI cross section of 3$\times$10$^{-49}$\,cm$^2$ at 40\,GeV/c$^2$ mass.  The science potential of a large, dual-phase xenon detector is detailed in~\cite{Aalbers:2022dzr}.  To address challenges related to the construction of next-generation Xe TPCs,  large-scale demonstrators have been built and are in operation, including a 2.6\, tall TPC~\cite{Baudis:2021ipf,Baudis:2023ywo,Brown:2023vgf}.  PandaX-xT is the next step in the PandaX programme at CJPL, with 43\,t of LXe target in the TPC (47\,t of LXe in total).  The experiment aims for a 200\,t\,y exposure for WIMPs, and, similar to DARWIN/XLZD, for a broad science reach~\cite{Wang:2023wrr}.

Detectors using superfluid $^4$He, targeted at light DM, are under development with HeRALD~\cite{SPICE:2023aqd}  and DELight~\cite{vonKrosigk:2022vnf} operating small prototypes. The two detectors are instrumented with transition edge sensors and magnetic micro-calorimeters, respectively, to observe both atomic signals, such as scintillation light, and quasiparticle (phonon and roton) excitations. The goal is to probe DM masses down to $\sim$300\,MeV, depending on the achieved energy thresholds.

\subsection{Bubble chambers}

To probe spin-dependent WIMP-nucleon interactions,  target  nuclei with uneven total angular momentum are required. A  favourable candidate 
is $^{19}$F, the spin of which is carried mostly by the unpaired proton.  Fluorine is part of the target of experiments using superheated liquids, such as 
the ones operated by the PICO collaboration~\cite{Krauss:2020ofg}. The bubble chambers are filled with superheated octafluoropropane (C$_3$F$_8$), with both acoustic and visual readout, allowing for an impressive rejection of ERs. A search in the PICO-60 C$_3$F$_6$ detector at SNOLAB, with an 
exposure of 1404\,kg\,d and an energy threshold of 2.45\,keV, yielded the most 
stringent constraint on the DM-proton SD cross section at 3.2$\times$10$^{-41}$cm$^2$ 
for a 25\,GeV particle mass.  PICO-40L started a physics run in a new design, which provides 2\,mm spatial position resolution and thus improved separation between single and multiple interactions in the chamber.  PICO-500, a detector which will use 250\,l of C$_3$F$_8$, is in construction in the Cube Hall of SNOLAB~\cite{Garcia-Viltres:2021swf}.

\subsection{Sodium Iodide experiments}

Large DM experiments using high-purity NaI(Tl) crystals have been constructed and are taking data to test the annual modulation signal reported by the DAMA/LIBRA collaboration. These observe the scintillation light produced when a particle interacts in a crystal.
The DAMA/LIBRA experiment~\cite{Bernabei:2023awy} at LNGS operates  250\,kg high-purity NaI(Tl) crystals, for a exposure of 1.33\,t\,y with an energy threshold 
of 1\,keV.  So far, it is the only experiment in the field that reported an annually modulated event 
rate. The statistical significance is  12.9\,$\sigma$ C.L. (20 annual cycles), with a modulation amplitude around 0.02\,events/(kg\,d\,keV) in the energy 
region (1-4)\,keV. These findings were interpreted as due to DM interactions via nuclear or electronic  recoils. 

The ANAIS-112 experiment~\cite{Coarasa:2024xec} in the Canfranc Underground Laboratory (LSC)  operates 112.5\,kg of NaI(Tl)  scintillators with an energy threshold of 1\,keV. A blind analysis of 3\,y  of data, for an exposure of 314\,kg\,y is consistent with an absence of  modulation and inconsistent with DAMA/LIBRA at 3\,$\sigma$ C.L.  The experiment has the potential to reach a 5\,$\sigma$ level after 8 years of data. The COSINE-100 experiment~\cite{Lee:2024bfl}, which concluded in March 2023 at the Yangyang  Underground Laboratory, operated 106\,kg of NaI(Tl) crystals in  a liquid scintillator,  with an energy threshold of 2\,keV. Results from a 97.7\,kg\,y exposure in the (2-6)\,keV energy range are consistent, at 68.3\%C.L., with both  the null hypothesis and DAMA/LIBRA's best fit value in the same energy range. An unambiguous conclusion was not reached due to the background level of the NaI (Tl) crystals. COSINE-200 will operate a higher mass experiment with reduced backgrounds, based on an in-house technology for crystal growth, and improved light yield. The experiment is currently in preparation stage at Yemilab, a new underground laboratory in South Korea. The SABRE experiment~\cite{McNamara:2024ncp} will operate  50\,kg of NaI(Tl) crystals, focussing on reaching a background level of 0.1\,events/(kg\,d\,keV), an order of magnitude below DAMA/LIBRA.  Twin detectors are being installed at LNGS and at the Stawell Underground Physics Laboratory in Australia.  

The COSINUS experiment, the construction of which was recently completed at LNGS, will operate  NaI cryogenic scintillating bolometers with 
both phonon and light readout~\cite{Hughes:2024nub}. Both signals will be measured with TESs,  allowing one to distinguish between ERs and NRs. The experimental facility houses a dry dilution refrigerator in a drywell inside a large water tank, which serves a muon veto and passive shield against radioactivity. Should a modulation signal be observed, COSINUS will be able to elucidate whether it is due to interactions with nuclei or with electrons.

\subsection{Ionisation detectors}

Silicon charged-coupled devices (CCDs) look for  low-energy ionisation events induced  in bulk silicon of high-resistivity. Charge 
resolutions around 1-2\,e$^-$ and extremely low leakage currents, at the level of  
few e$^-$\,mm$^{-2}$d$^{-1}$ can be reached. The interacting particle is identified based on the recorded track pattern. 
The DAMIC and SENSEI experiments~\cite{Traina:2024ics} at SNOLAB yielded leading constraints on DM-electron scattering.  The DAMIC-M experiment~\cite{Privitera:2024tpq} at LSM plans for a kg-size detector mass by the end of 2024. The goal is to achieve an energy threshold of 2-3 electrons and to probe the DM-nucleon cross section down to few$\times$10$^{-43}$cm$^2$  around 2-3\,GeV and the DM-electron cross section down to 2$\times$10$^{-41}$cm$^2$ at 10\,MeV mass. Oscura is a planned next-generation detector with 10\,kg of active Si mass.
 
Germanium ionisation detectors operated at 77\,K can reach sub-keV energy thresholds and very low background levels. While they cannot distinguish between ERs and NRs, pulse shape discrimination is employed to discriminate between bulk  and surface events with incomplete charge collection~\cite{Zhang:2024asp}.  The CDEX-10 experiment  at CJPL  uses p-type, point-contact Ge detectors operated in  liquid nitrogen to probe DM masses down to 3\,GeV~\cite{CDEX:2019isc}.

The NEWS-G collaboration operates spherical proportional counters (SPCs) with light, gaseous targets (H, He, Ne) optimised for sensitivity to sub-GeV DM particles. SPCs feature high gains and low intrinsic electronic noise, allowing one to detect single ionisation electrons and thus to reach low energy thresholds~\cite{NEWS-G:2022kon}.  Pulse shape discrimination is used to reject surface events. NEWS-G recently developed a 140\,cm diameter, high-purity  Cu detector, with a multi-anode sensors readout. The detector is shielded by layers of lead, with an inner 3\,cm thick layer of Roman lead, placed inside a neutron shield made of  high density polyethylene.  After a commissioning run at LSM, the detector was moved to SNOLAB where it currently acquires data~\cite{Coquillat:2024tfa}. 
 
The TREX-DM detector is a high-pressure TPC with mikrobulk Micromegas readout, installed at LSC~\cite{Castel:2019ngt}.  The TPC can be filled with Ar or Ne mixtures to search for DM particles in the mass range from $\sim$10\,GeV to sub-GeV, depending on the achieved energy threshold (0.9\,keV to 50\,eV for ERs). The threshold depends on the gain of the Micromegas, and, to increase the gain, a new electron pre-amplification stage with a Gas Electron Multiplier (GEM) on top of the Micromegas was recently adopted.  The experiment currently acquires science data at LSC.

\subsection{Directional detectors}

Directional experiments aim to measure the recoil direction, which is correlated to the galactic motion towards constellation Cygnus. The recoil spectrum would be peaked in the opposite direction and, ideally, a detector should measure the axis and sense of a DM-induced recoil. A signature would require only a few tens of events and terrestrial, seasonal modulations would be unable to fake a DM signal~\cite{Mayet:2016zxu}. The main challenge is to achieve a good angular resolution and to distinguish between head and tail at low recoil energies.

Several directional detectors are in operation and  Cygnus is a proto-collaboration to coordinate the R\&D efforts for gas based TPCs with 1\,keV threshold. CYGNO~\cite{Amaro:2023dxb} aims to built a TPC with 50\,cm drift, filled with a He:CF$_4$ gas mixture at room temperature and atmospheric pressure at LNGS. The liberated charge in a particle interaction is drifted towards an amplification stage consisting of a triple GEM structure, where the charge is multiplied and also light is produced. The readout consists of light detectors, PMTs and scientific CMOS cameras, yielding the energy and z-position of events and the two-dimensional track projections, respectively. The combined information thus allows for a three-dimensional track reconstruction. 

Other techniques for directional detectors are based  on fine-grained nuclear emulsions (solid-state detectors with  silver halide crystals uniformly dispersed in a gelatine film, where each crystal works  as a sensor for charged particles), as proposed by NEWSdm~\cite{DiCrescenzo:2023xhl}. These act as target and nanometric tracking device, with the expected NR tracks being sub-$\mu$m in size. Due to the small crystal size and larger number density, a superior spatial resolution compared to gaseous detectors is obtained.  For a recent review on directional recoil detection I refer to Ref.~\cite{Vahsen:2021gnb}.

\subsection{Other technologies}

To search for DM particles with masses below $\sim$1\,GeV many new technologies were proposed and are currently being explored. These are based on materials with small band gaps for electron excitations, such  as superconductors, Dirac materials, graphene, superfluid helium, and polar crystals, to name a few~\cite{Essig:2016crl, Hochberg:2017wce, Knapen:2017ekk}.  One approach is to use superconducting nanowires (SNSPDs) as both target and sensor, with 
first bounds on DM-electron interactions obtained with a 4.3\,ng WSi prototype~\cite{Hochberg:2019cyy}.  This approach is used by Qrocodile (Quantum Resolution-Optimized Cryogenic Observatory for Dark matter Incident at Low Energy), a new interdisciplinary collaboration between condensed matter, astroparticle experiment, quantum sensing and particle theory~\cite{QROCODILE:2024nqm}. The goals are to scale-up the SNSPDs detector mass to areas  of $\sim$cm$^2$ and beyond, to extend the energy threshold to the
fundamental limit, as well as to understand and reduce the background sources.  Overall, the detection of light DM via collective excitations in condensed matter systems is a rapidly growing area of research, and I refer to ~\cite{Lin:2022hnt,Essig:2022dfa,Mitridate:2022tnv} for recent overviews.

\section{Conclusions}

Direct DM detection experiments must be sensitive to an enormous range of particle masses and cross sections. This is reflected in the diverse range of employed technologies and target materials. The principal experimental challenges  are to lower the energy thresholds further, to reduce and characterise the background sources and at the same time to increase the DM target masses. The main goals are to discover a new, dark particle species and to explore the experimentally accessible parameter space, until the observed recoil rates will be dominated by astrophysical neutrinos. 

\bibliographystyle{iopart-num} 
\bibliography{baudis_dm_paris}

\end{document}